\documentclass[twoside]{article}
\usepackage{fleqn,espcrc2,rotating}

\usepackage{graphicx}
\input{psfig}
\usepackage{epsfig}

\newcommand{\AmS}{{\protect\the\textfont2
  A\kern-.1667em\lower.5ex\hbox{M}\kern-.125emS}}

\title{
\begin{flushright}
Fermilab-Conf-00/229\\Sept2000
\end{flushright}
Charged Kaons at the Main Injector (CKM)}

\author{
\small
J. Frank, S. Kettell, R. Strand \\
\scriptsize
Brookhaven National Laboratory, Upton, NY, USA \\
\small
\vspace*{0.1in}
\small
L. Bellantoni, R. Coleman, P.S. Cooper, M. Crisler,
C. Milste\'ne$^{\dagger}$,H. Nguyen, E. Ramberg, R. S. Tschirhart \\
\scriptsize
Fermi National Accelerator Laboratory, Batavia, IL, USA \\
\small
\vspace*{0.1in}
\small
G.V.~Britvich, V.~Kubarovsky, L.~Landsberg,V.~Molchanov,
V.~Obraztsov,  S.V.~Petrenko, V.~Polyakov, D.~Vavilov, O.~Yushchenko \\
\scriptsize
Institute of High Energy Physics, Serpukhov, Russia \\
\small
\vspace*{0.1in}
\small
J. Engelfried, A. Morelos \\
\scriptsize
Instituto de Fisica, Universidad Autonoma de San Luis Potosi, Mexico \\
\small
\vspace*{0.1in}
\small
M. Campbell \\
\scriptsize
University of Michigan, Ann Arbor, Michigan 48109 \\
\vspace*{0.1in}
\small
K. Lang \\
\scriptsize
University of Texas at Austin, Austin, Texas 78712 \\
\vspace*{0.1in}
\small
C. Dukes, K. Nelson \\
\scriptsize
University of Virgina, Charlottesville, VA 22901 \\
\small
\vspace*{0.2in}
Comments: 5pages, 7 figures Presented by C. Milste\'ne {Fermi National Laboratory-For the E905 
Collaboration,Proceeding of the IV International Conference on Hyperons, Charm and Beauty Hadrons, 
Valencia, Spain - June 27-July 1-2000. to be published in Nucl.Phys.B\\ }}

\begin{document}

\begin{abstract}

 The CKM collaboration is proposing to measure the branching ratio of the rare 
K decay $K^+ \to \pi^+ \nu \bar{\nu}$ at the Main Injector at Fermilab. 
Our goal is to be able to observe $\simeq$ 100 events, for a Standard Model
branching ratio of $\simeq 1.\ 10^{-10}$. This implies that we must be able 
to reduce the background to a few events at a reasonable cost.
\vspace{1pc}
\end{abstract}

\maketitle

\section{The Physics}

\begin{figure}[htb]
\vspace{9pt}
\centerline{
  \epsfig{figure= 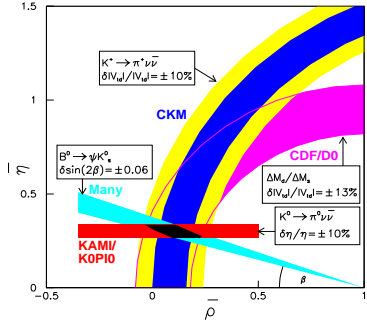,width=65mm,height=50mm}
}
\caption { Experiments proposed sensitivity in the measurement of the CP 
violating parameters of the Standard Model .}
\label{PHYS}
\end{figure}

 The branching ratio for the  decay $K^+ \to \pi^+ \nu \bar{\nu}$ was first 
 calculated by Inami and Lin \cite{INAML} as an isospin rotation 
 from $K^+ \to \pi^0 e^+ \nu$. The measurement of the branching ratio 
 of  $K^+ \to \pi^+ \nu \bar{\nu}$ alone will allow one to obtain the 
magnitude of $V_{td}$, the element in the Cabbibo Kobayashi Maskawa matrix, 
which controls Standard Model CP violation. In conjunction with the neutral
channel decay $K^0 \to \pi^0 \nu \bar{\nu}$ and measurements in the B sector, 
this will provide an overconstrained measurement of both the real and the 
imaginary part of $V_{td}$. This will address the question on CP violation: 
Can it be completely accounted for by the Standard Model. Figure~\ref{PHYS} 
illustrates the proposed sensitivity of various experiments that will measure 
$\rho$ and $\eta$, the CP violating parameters of the standard model.

 The Standard Model predicts a branching ratio for the decay of 
$[0.8\pm 0.3]\bullet 10^{-10}$ in Buras's unitarity triangle \cite{BURAS}.
 One event has been observed by the BNL experiment E787 in a stopped 
$K^+$ experiment. We propose \cite{PROPOS} to 
measure the decay in flight of $K^+ \to \pi^+ \nu \bar{\nu}$ and obtain 
$>$70 events in 2 years of data taking .  This will meet the precision suggested 
by the theory. An intermediate stage, the observation of 5-10 events, is the 
goal of an approved BNL experiment E949, which is the continuation of E787. 
Some members of the CKM and E949 are now working in collaboration with each 
other as two stages of a physics program.

\section{The Background}

 One of the challenges of the experiment is to control the background which
is 10 orders of magnitude larger than the expected signal.
\subsection{ The $K^+ \to \pi^+ \pi^0$ background}

 This decay, with branching ratio $\simeq$ 21\%, can mimic the signal
if the two $\gamma$ from the decay of the $\pi^0$ are undetected. It will be 
controlled  by a combination of kinematic rejection and a photon veto system.
The CKM detector must veto photons with a very high efficiency.
\subsection { The $K^+ \to \mu \nu$ background}

 In this decay , with a branching ratio $\simeq$63.5\%, the $\mu$ can be 
misidentified as a $\pi$ .This is controlled by a muon veto and 
RICH systems.
\subsection { Background from Interactions}

The interactions of the undecayed beam  or daughter $\pi$s contribute to the 
background. In order to minimize this contribution, as little material 
as possible should be encountered by the particles. This dictates the choice
of the various detector elements.

\section{The Apparatus}
\begin{figure}[htb]
\vspace{9pt}
\centerline{
  \epsfig{figure= 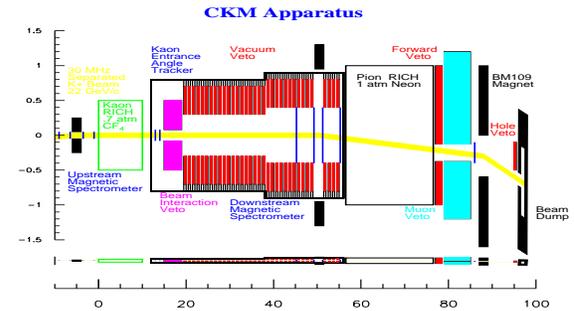,width=80mm,height=51mm}
}
\caption { The Detector }
\label{APP}
\end{figure}

 In Figure~\ref{APP} are shown the detector elements which will be 
detailed below. To detect $K^+$ hadronic interactions before the decay volume 
entrance, a hadron calorimeter surrounding the beam (not described)is used .
\subsection{The RICH and the Spectrometers}

 The charged particles in the reaction, namely the K and the $\pi$ 
are both measured by Ring Imaging CHerenkov counters (RICH) as velocity 
vectors and as momentum vectors in upstream and downstream spectrometers. 
 The gas in the pion RICH is chosen in order to meet the low dispersion 
required for good resolution. In order to allow as little material in 
the beam as possible rather thin mirrors and windows are mandatory. This
require the RICHs to operate at or near atmospheric pressure. 
These considerations have contributed to the choice of the K beam to 
be 22GeV energy and the radiator gas of the decay pion RICH to be neon.
 The $\pi$ momentum resolution achieved by the RICH is better than 1\% for 
this configuration. The K and $\pi$ magnetic spectrometers use conventional
trackers, silicon strips and straw tubes respectively.
 Straws have been chosen to minimize the material in the beam.
They are 5mm in diameter with a 20 $\mu$m sense wire, 5 to 8 straw layers
per station (x,y,u,v). $4\bullet 10^-4$ $X_0$, with 150$\mu$m hit resolution 
per layer. The straws are located in the vacuum decay volume.
\subsection{The $\gamma$ veto system}

 As already mentioned one of the main tasks of the experiment is to control 
 the background from the $K^+ \to \pi^+ \pi^0$ decay. This points to the 
extreme importance of the $\gamma$ veto system.
 The veto system has a component surrounding the decay volume, where
the decay takes place and a forward component, after the $\pi$ 
RICH to deal with high energy $\gamma$'s. We want to minimize  interactions
which are an important component of the non-Gaussian tails of the background. 
Therefore, the component surrounding the decay volume will be located inside
the vacuum vessel with a vacuum of $\simeq 10^{-6}$ torr and we call this 
Vacuum Veto System (VVS). The VVS is made of a sandwich of plastic scintillator 
and lead 5mm/1mm each, stacked in modules 50 cm long and 40cm radially. 
There are over 80 such sandwich layers per module and 34 such 
modules in the VVS. There is a void of 50cm between each module which 
corresponds to an angle of incidence of $>40$ degrees for $\gamma$. This
angle is not reached by the photons coming from the 
$K^+ \to \pi^+ \pi^0 \to\gamma \gamma$ decay kinematics . This allows us 
to reduce the volume and the cost by a factor two.

\begin{figure}[htb]
\vspace{9pt}
\centerline{
  \epsfig{figure= 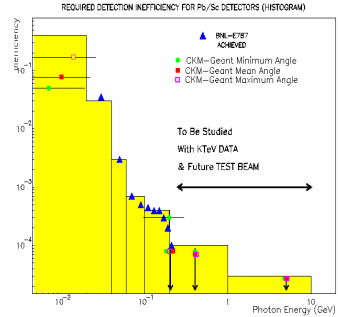,width=49mm,height=55mm,angle=-90}
} 
\caption { VVS Required inefficiencies. The triangles were measurements
achieved by E787.The circles represent the inefficiencies
reached by Geant simulations studies at low and high energies.
 }
\label{SPEC}
\end{figure}

 The required VVS detection inefficiency is given in Figure~\ref{SPEC}.
In the region of medium energies the requirements have been achieved by
E787 with a comparable veto system. 
 
 For the Forward Veto System (FVS) a choice of various techniques 
are avalaible, and are still under study. We are assuming, for the moment,
a setup of Sc/Pb 5mm/1mm  sandwiches for this calorimeter as well.
\subsection{The Muon Veto}

 Another important component of the background is the $K^+ \to \mu^+ \nu$ 
 which has a branching ratio of $\simeq$63.5\%. A muon veto follows the FVS.
 The non-interacting beam goes through a hole in the FVS and muon
veto system and is dumped afterward by a magnetic field into the beam dump.
\section{The Beam}

\begin{figure}[htb]
\vspace{9pt}
\centerline{
  \epsfig{figure= 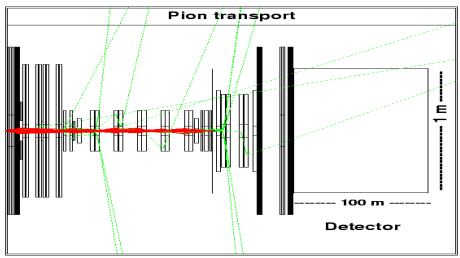,width=66mm,height=47mm}
}
\caption { The behavior of the $\pi$ part of the beam. }
\label{PSTOP}
\end{figure}

\begin{figure}[htb]
\vspace{9pt}
\centerline{
  \epsfig{figure= 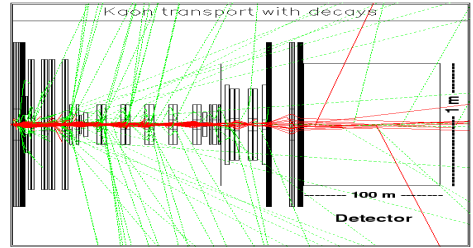,width=67mm,height=46mm}
}
\caption { The behavior of K part of the beam }
\label{KDET}
\end{figure}

\begin{figure}[htb]
\vspace{9pt}
\centerline{
  \epsfig{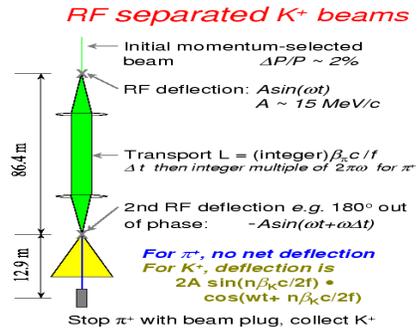}
}
\caption { The behavior of the $\pi$ part of the beam }
\label{RF}
\end{figure}

A very important component of our experiment is the beam. In order to get 
$\simeq$ 6MHz of $K^+$ decays in the fiducial volume we need a $\simeq$30 MHz 
$K^+$ beam with a ratio of $K/\pi$ of$>2$, whereas nature tends to provide far
more $\pi$'s than K's, shown Figure~\ref{PSTOP} and Figure~\ref{KDET}. 
We will construct a new RF separated beam based upon 
3.9 GHz transverse mode superconducting RF cavities, being developed at 
Fermilab for this purpose.  
 We use a set of 2 RF cavities properly spaced from each other in order for 
the $\pi$s to be deflected by the 1st cavity, and then deflected back by the 
same amount, by the 2nd cavity. The $\pi$s are absorbed by a beam stopper. 
With the cavities phased this way, protons which arrive at the second cavity 
just one RF period behind the $\pi$s are also undeflected. Ks, however, 
arrive $180^\circ$ in RF phase behind
the $\pi$ and are deflected again with the same sign, causing most of them to 
miss the beam stopper. A cartoon of the operation is shown in Figure~\ref{RF}.

 In Figure~\ref{PSTOP} we show the behavior of the $\pi$ component of the beam
which after having been deflected back by the second cavity is stopped. 
Shown in the same figure are the $\mu$s coming from $\pi$ decays in the beam 
line.

 In Figure~\ref{KDET} the K component is shown with the decay $\mu$s. The 
Ks which avoid the stopper are shown reaching the detector.

\section{Status}
 We are in the process of building prototypes for the photon veto system,
the straw tubes and the super-conducting RF cavities. Simulations are being 
done in parallel for both the straws and the veto systems as well as for 
the beam line design.

\section{Conclusion}

\begin{figure}[htb]
\vspace{9pt}
\centerline{
  \epsfig{figure= 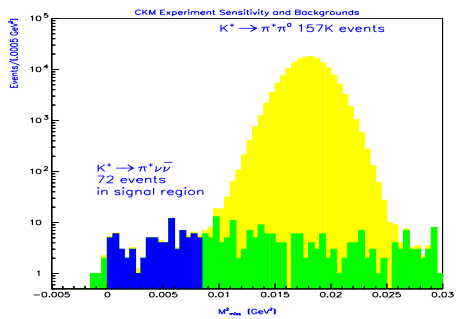,width=65mm,height=45mm}
}
\caption { The Missing Mass Square of the generated signal and backgrounds
corresponding to the whole experiment }
\label{MMSQ}
\end{figure}
 
Figure~\ref{MMSQ} shows the expected signal and $K^+ \to \pi^+ \pi^0$ background 
after the photon vetos have been applied from  two years of running as 
a function of the missing mass squared of the neutral system recoiling against
the $\pi^+$. 72 events survive in a signal region with 3 unvetoed and 
mismeasured $K^+ \to \pi^+ \pi^0$ event. The total background estimate for
this sample, including $K^+ \to \mu^+ \nu$ , interactions and accidentals is 
$< 8$ events. Work is now underway to refine and improve our proposal with 
the intention of seeking approval in the spring 2001.

\end{document}